\documentclass{PoS}

\usepackage{amsxtra}
\usepackage{cite}

\newcommand \VEV [1] {\left\langle{#1}\right\rangle}

\newcommand{\beq}{\begin{eqnarray}}
\newcommand{\eeq}{\end{eqnarray}}

\newcommand{\gev}{{\rm GeV}}
\newcommand{\mev}{{\rm MeV}}

\newcommand{\be}{\begin{equation}}
\newcommand{\ee}{\end{equation}}
\newcommand{\lwrsim}{\raise0.3ex\hbox{$<$\kern-0.75em\raise-1.1ex\hbox{$\sim$}}}
\newcommand{\lgrsim}{\raise0.3ex\hbox{$>$\kern-0.75em\raise-1.1ex\hbox{$\sim$}}}

\newcommand{\msbar}{\overline{\rm MS}}
\newcommand \vev [1] {\langle{#1}\rangle}

\title{$\Lambda_{QCD}$ from gluon and ghost propagators}

\ShortTitle{$\Lambda_{QCD}$ from gluon and ghost propagators}

\author{F. De soto\\
        Dpto. Sistemas F\'isicos, Qu\'imicos y Naturales, \\ 
        Universidad Pablo de Olavide, 41013 Sevilla, Spain.\\
        E-mail: \email{fcsotbor@upo.es}}

\author{\speaker{M. Gravina}, O. P\`ene\\
        Laboratoire de Physique Th\'eorique\footnote{Unit\'e Mixte 
        de Recherche 8627 du Centre National de
        la Recherche Scientifique}\\
        {Universit\'e de Paris-Sud XI, B\^atiment 210, 91405 Orsay Cedex,
        France}\\
        E-mail: \email{Mario.Gravina,Olivier.Pene@th.u-psud.fr}}

\author{J. Rodr\'iguez-Quintero\\
        Dpto. F\'isica Aplicada, Fac. Ciencias Experimentales,\\
        Universidad de Huelva, 21071 Huelva, Spain\\
        E-mail: \email{jose.rodriguez@dfaie.uhu.es}}

\abstract{
\begin{center}
\raisebox{5mm}{\hbox{
\includegraphics[width=3cm]{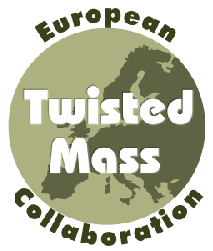}}}
\end{center}

Fundamental quantities of QCD, such as the strong coupling 
and $\Lambda_{QCD}$, are studied in the framework of lattice QCD with
$N_f=2$ twisted mass fermions. In particular, the contact between
lattice and continuous calculations is made by comparing the
renormalized ghost-gluon vertex in MOM scheme with 4-loop
perturbative results. A power correction is needed in order to have agreement 
between the two descriptions. This suggests the presence of a dimension-two 
$\VEV{A^2}$ gluon condensate whose value is found to be higher than 
in the quanched case.
}
\FullConference{The XXVII International Symposium on Lattice Field Theory - LAT2009\\
		 July 26-31 2009\\
		 Peking University, Beijing, China}

\begin{document}

\section{Introduction}

The lattice provides a very elegant way of calculating renormalized 
observables. In this framework several methods are known to extract the running 
of the QCD coupling constant, which allows for the determination of the 
QCD scale, $\Lambda_{QCD}$ and for the study of infrared properties.
In the case of quenched world the mismatch between 
the perturbative running and the lattice one 
has revealed the presence of a non-null gluon condensate
of dimension two that, being non-gauge invariant, has motivated 
the research of its possible implications for the gauge-invariant world.

In this note we apply the already established methods for
$N_f=2$ dynamical quarks, including light up and down quarks. 
$N_f=2+1+1$ lattice simulations are already being performed, 
thus a realistic lattice estimate of $\Lambda_{\overline{MS}}$ directly
comparable with experimental results will become inmediatly accesible.

In particular here we focus on the study of the ghost-gluon vertex in the 
configuration of vanishing incoming ghost-momentum. Only in this case 
the ghost-gluon vertex can be related directly to the bare and 
ghost propagators, making calculations simpler. 

\section{Taylor scheme}

\subsection{Definitions}

In~\cite{Boucaud:2008gn} was shown that the so-called Taylor scheme is
the only one where the coupling can be cumputed from two-point
Green functions, due to Taylor's theorem.
We write Landau gauge gluon and ghost propagators as:
\beq
\left( G^{(2)} \right)_{\mu \nu}^{a b}(p^2,\Lambda) &=& \frac{G(p^2,\Lambda)}{p^2} \ \delta_{a b}
\left( \delta_{\mu \nu}-\frac{p_\mu p_\nu}{p^2} \right) \ ,
%\langle \widetilde{A^a_\mu}(-p) \widetilde{A^b_\nu}(p) \rangle \ ,
\nonumber \\
\left(F^{(2)} \right)^{a,b}(p^2,\Lambda) &=& - \delta_{a b} \ \frac{F(p^2,\Lambda)}{p^2} \ ;
\eeq
with $\Lambda$ the regularisation cutoff. The renormalized dressing functions,
$G_R$ and $F_R$ are defined through :
\beq
G_R(p^2,\mu^2)\ &=& \ \lim_{\Lambda \to \infty} Z_3^{-1}(\mu^2,\Lambda) \ G(p^2,\Lambda)\nonumber\\
F_R(p^2,\mu^2)\ &=& \ \lim_{\Lambda \to \infty} \widetilde{Z}_3^{-1}(\mu^2,\Lambda)\ F(p^2,\Lambda) \ ,
\eeq
with MOM renormalization condition
\beq\label{bar2}
G_R(\mu^2,\mu^2)=F_R(\mu^2,\mu^2)=1 \ .
\eeq
Due to Taylor's non-renormalization theorem, the renormalized
coupling defined from the ghost-gluon vertex {\bf with a zero
incoming ghost momentum} can be computed from ghost and gluon
propagators using:
\begin{equation} \label{alphaT}
\alpha_T(\mu^2) \equiv \frac{g^2_T(\mu^2)}{4 \pi}=  \
\lim_{\Lambda \to \infty} \frac{g_0^2(\Lambda^2)}{4 \pi}
G(\mu^2,\Lambda^2) F^{2}(\mu^2,\Lambda^2) \ ;
\end{equation}
what has been called Taylor~\footnote{From now on, the quantities
expressed in this scheme will carry the $T$ index.}
scheme~\cite{Boucaud:2008gn}

\subsection{Perturbation theory and OPE}

The perturbative running of $\alpha_T$ is known up to four loops~\cite{Chetyrkin:2004mf},
\beq
  \label{betainvert}
%  \begin{split}
      \alpha_T(\mu^2) &=& \frac{4 \pi}{\beta_{0}t}
      \left(1 - \frac{\beta_{1}}{\beta_{0}^{2}}\frac{\log(t)}{t}
     + \frac{\beta_{1}^{2}}{\beta_{0}^{4}}
       \frac{1}{t^{2}}\left(\left(\log(t)-\frac{1}{2}\right)^{2}
     + \frac{\widetilde{\beta}_{2}\beta_{0}}{\beta_{1}^{2}}-\frac{5}{4}\right)\right) \\
     + \frac{1}{(\beta_{0}t)^{4}}&&
\left(\frac{\widetilde{\beta}_{3}}{2\beta_{0}}+
   \frac{1}{2}\left(\frac{\beta_{1}}{\beta_{0}}\right)^{3}
   \left(-2\log^{3}(t)+5\log^{2}(t)+
\left(4-6\frac{\widetilde{\beta}_{2}\beta_{0}}{\beta_{1}^{2}}\right)\log(t)-1\right)\right)
%   \end{split}
\eeq
with $t=\ln \frac{\mu^2}{\Lambda_T^2}$ and the perturbative coefficients:
\beq\label{betacoefs}
\widetilde{\beta}_0 &=& \overline{\beta}_0 = 11 - \frac 2 3 N_f
\nonumber \\
\widetilde{\beta}_1 &=& \overline{\beta}_1 = 102 - \frac{38} 3 N_f
\nonumber \\
\widetilde{\beta}_2 &=& \overline{\beta}_2 -\overline{\beta}_1 c_1 + \overline{\beta}_0 (c_2-c_1^2)
\nonumber \\
&=&3040.48 \ - \ 625.387 \ N_f \ + \ 19.3833 \ N_f^2
\nonumber \\
\widetilde{\beta}_3 &=& \overline{\beta}_3 - 2 \overline{\beta}_2 c_1 + \overline{\beta}_1 c_1^2
+ \overline{\beta}_0 (2 \ c_3 - 6 \ c_2 c_1 + 4 \ c_1^3)
\nonumber \\
&=&  100541 \ - \ 24423.3 \ N_f \ + \ 1625.4 \ N_f^2 \ - \ 27.493 \ N_f^3
\ ,
\eeq

The parameters $\Lambda_{QCD}$ in two schemes can be perturbatively related
at high energy. In particular, from the $T$-scheme to $\overline{MS}$ this relationship
reads:
\beq\label{ratTMS}
\frac{\Lambda_{\overline{\rm MS}}}{\Lambda_T} \ = \ e^{\displaystyle -\frac{c_1}{2 \beta_0}} \ = \
e^{\displaystyle - \frac{507-40 N_f}{792 - 48 N_f}}
\ .
\eeq

Following the Operatore Product Expansion (OPE) program both ghost 
and gluon propagators
show the appearance of a non-perturbative power correction driven
by the non-gauge invariant dimension-two gluon condensate 
(see \cite{Boucaud:2008gn}, \cite{Boucaud:2001st} and referencies therein).
Including power corrections at tree-level in ghost and gluon dressing 
functions, one can rewrite (\ref{alphaT}) as

\beq\label{alphaNP}
\alpha_T(\mu^2) &=& \alpha^{\rm pert}_T(\mu^2) \left(
 1 + \frac{9}{\mu^2} \frac{g^2_T(q_0^2) \langle A^2 \rangle_{R,q_0^2}} {4 (N_C^2-1)}
\right) \ ,
\eeq
where $q_0^2 \gg \Lambda_{\rm QCD}$ is some perturbative scale and
the running of the perturbative part is described by 
equation (\ref{betainvert}).
This formula will be used for the data analysis in the next section that does depend
on two parameters, $\Lambda_{QCD}$ and $\langle A^2\rangle$, that will be fitted.

\section{Lattice setup and role of $H(4)$  orbits}

The results presented here are based on the gauge field configurations generated by the
European Twisted Mass Collaboration (ETMC) with the tree-level improved Symanzik
gauge action~\cite{Weisz82}  and the twisted mass fermionic action~\cite{Frezzotti:2000nk}  at
maximal twist, discussed in detail in refs.~\cite{Boucaud:2007uk}-~\cite{Dimopoulos:2008sy}.

We preliminarly exploited  100 ETMC gauge configurations obtained
for $\beta=3.9$ ($\mu=0.0085$), 60 for $\beta=4.05$ ($\mu=0.006$)
and 100 for $\beta=4.2$ ($\mu=0.002$) simulated on $24^3\times 48$
lattices, corresponding to $N_f=2$ in order to compute the gauge-fixed 2-point gluon and ghost Green functions.

For fixing Landau gauge in the lattice we minimise the functional
\beq
F_U[g] = Re \sum_x \sum_\mu \left(1-\frac{1}{N}g(x)U_\mu(x)g(x+\mu) \right)
\eeq
respect to the gauge transform $g$.
Ghost propagator is computed in Landau gauge as the inverse of the Faddeev-Popov operator, that
is written as the lattice divergence,
\beq
M(U) = -\frac{1}{N} \nabla \cdot \widetilde{D}(U)
\eeq
where the operator $\widetilde{D}$ acting on an arbitrary element of the Lie algebra, $\eta$ reads:
\beq
\widetilde{D}(U) \eta(x) = \frac{1}{2} \left(U_{\mu}(x)\eta(x+\mu) -
\eta(x)U_\mu(x)+\eta(x+\mu)U_\mu^\dagger-u_\mu^\dagger(x)\eta(x)\right)
\ . \eeq
More details on the lattice procedure for the inversion of Faddeev-Popov operator can be found
on~\cite{Boucaud:2005gg}.

As we intend to fit the running of $\alpha_s$, our interest is to have, on one hand the
highest momenta accesible and, on the other the highest number of data points to perform
the fit. When working at a given lattice spacing, the momentum window has to be limited
due to the presence of high discretization errors. These lattice artifacts are due to the
breaking of the  rotational
symmetry of the euclidean space-time when using an hypercubic lattice, where this symmetry
is restricted to the discrete H(4) isometry group. These artifacts can be
illustrated as the difference between the lattice momenta,
\beq
\tilde{p}_\mu = \frac{1}{a} \sin ap_\mu
\eeq
and the continuum ones,
\beq
{p}_\mu = \frac{2\pi n}{N a} \qquad n=0,1,\cdots,N \ .
\eeq
Clearly these two momenta will differ except in the limit $n/N\to 0$.
Following what was recently discussed in~\cite{Becirevic:1999uc} and 
\cite{deSoto:2007ht}, let us 
consider an adimensional lattice correlation function $Q$
that depends on the lattice momentum $a\tilde{p}_\mu$
and some mass scale $a\Lambda$: $Q \equiv Q(a^2\tilde{p}^2,a^2\Lambda^2)$ .
The lattice momentum can be developed as:
\begin{equation}
a^2\tilde{p}_\mu^2 = a^2p_\mu^2 + c_1 a^4 p_\mu^4 + \cdots
\end{equation}
with $c_1$ a constant that depends on the discretization chosen.
Then:
\begin{eqnarray}
a^2\tilde{p}^2 \equiv \sum_{\mu=1}^{4} a^2\tilde{p}_\mu^2 = a^2p^2
+ c_1 a^4 p^{[4]} + \cdots = a^2p^2 \left( 1 + c_1 a^2
\frac{p^{[4]}}{p^2} + \cdots \right)
\end{eqnarray}
where $p^{[4]}=\sum_{\mu=1}^{4} p_\mu^4$.
If the lattice spacing is small, $\epsilon=a^2 p^{[4]}/p^2<<1$ and
we can develop $Q$ in powers of $\epsilon$:
\begin{eqnarray}
Q(a^2\tilde{p}_\mu^2,a^2\Lambda^2) &\equiv& Q\left(a^2p^2 \left( 1
+ c_1 a^2 \frac{p^{[4]}}{p^2} + \cdots \right),a^2\Lambda^2\right)\\
&=& Q(a^2p^2,a^2\Lambda^2) +
\left.\frac{dQ}{d\epsilon}\right|_{\epsilon=0} a^2
\frac{p^{[4]}}{p^2} + \cdots
\end{eqnarray}
$H(4)$ methods are based on the appearance of a $\mathcal{O}(a^2)$
corrections driven by a $p^{[4]}$ term. The basic method is to fit
between the whole set of orbits sharing the same $p^2$ the
coefficient $R$ and the extrapolated value of $Q$
free from H(4) artefacts. In particular we assumed that the coefficient 
\[
\left.R(a^2p^2,a^2\Lambda^2) = \frac{dQ\left(a^2p^2 \left( 1 + c_1 \epsilon + \cdots
\right),a^2\Lambda^2\right)}{d\epsilon}\right|_{\epsilon=0}
\]
has a smooth dependence on $a^2p^2$
over a given momentum window. This can be achieved by developing
$R$ as $R=R_0+R_1 a^2p^2$ and making a global fit in a momentum window
between $(p-\delta,p+\delta)$ to extract the extrapolated value of
$Q$ for the momentum $p$ and shifting the window for every lattice
momentum. This procedure of fitting is somehow different from the 
previous one, since the extrapolation does not rely on any 
particular assumption for the functional form of
$R$. On the other, the systematic error coming from the
extrapolation can be estimated by modifying the width of the
fitting window.

\section{Results}

\subsection{Calibration of lattice spacings}

The running of $\alpha_T$ given by the combination of Green
functions in eq. (\ref{alphaT}) does depend in principle on the
momentum and the cut-off. Nevertheless, if we are not far from the
continuum limit, and discretization errors are treated properly, 
the coupling will depend only on the
momentum (except, maybe, finite volume errors at low momenta).

The procedure to compute the ratio of lattice spacings is then
straightforward: it can be obtained by requiring the estimates of
$\alpha_T$ for two different simulations (two different $\beta$'s)
to match properly each other. This method has proven to be
successful in quenched lattice simulations~\cite{Boucaud:2008gn},
with a deviation with respect to usual Sommer parameter estimates lower
than $5\%$.

The $N_f=2$ results can be seen in figure \ref{fig:alpha_scaling}, where
the lattice spacing for the lower $\beta$ ($\beta=3.9$) has been
assumed to coincide with the one given in~\cite{Boucaud:2007uk} and the other
two, for $\beta=4.05$ and $\beta=4.20$ are fitted to match the
data.

\begin{table}
\label{tab:scaling} 
\begin{center}
\begin{tabular}{|c|c|c|c|}
\hline
& This paper & Sommer scale & deviation (\%))\\
\hline
$a(3.9)/a(4.05)$ & 1.223(3) & 1.277 & 4.2 \\
\hline
$a(3.9)/a(4.2)$ & 1.503(5)& 1.547 &  2.9 \\
\hline
\end{tabular}
\end{center}
\caption{\small Best-fit parameters for the ratios of
lattice spacings. The error is purely statistics.}
\end{table}

\begin{figure}[h!]
\begin{center}
\begin{tabular}{cc}
\includegraphics[width=6.5cm]{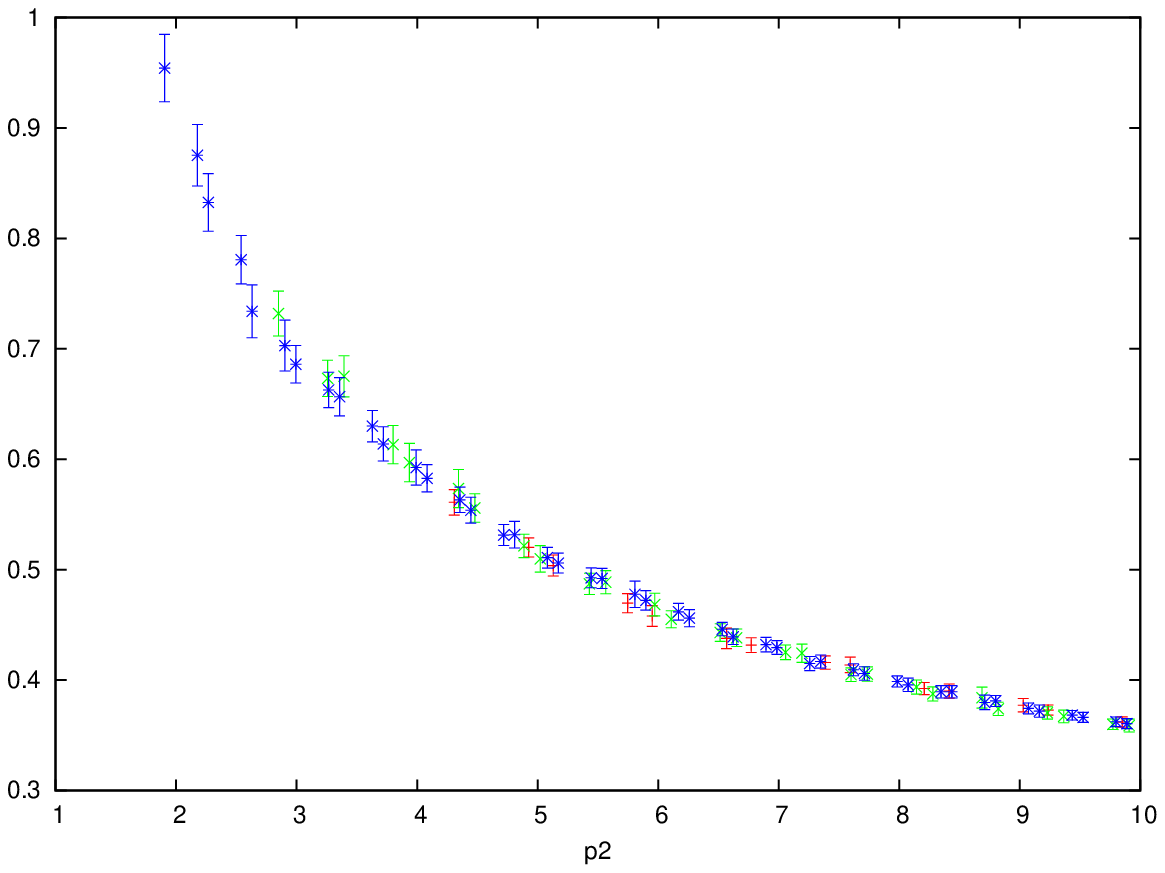} &
\includegraphics[width=6.5cm]{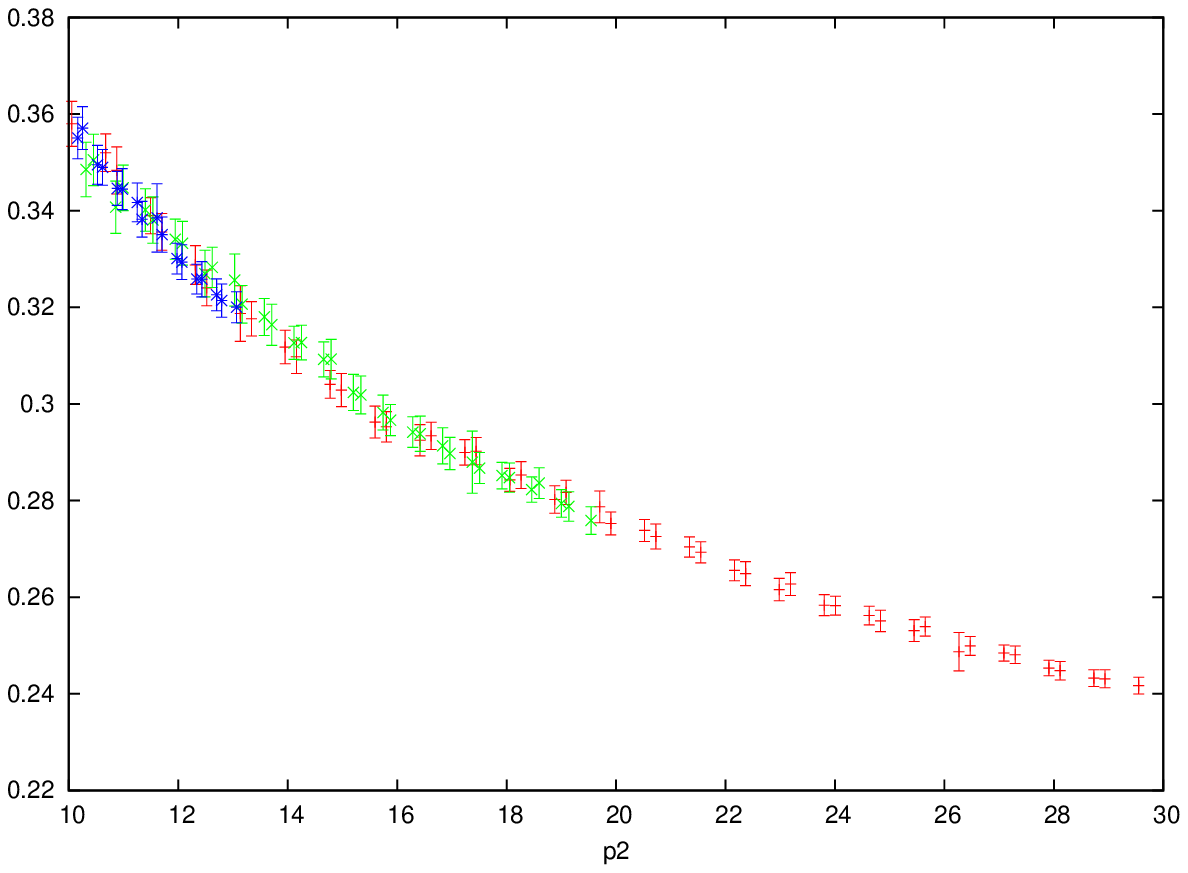}
\end{tabular}
\end{center}
\label{fig:alpha_scaling} 
\label{fig:alpha_scaling} \caption{\small QCD coupling defined by 
from the three lattice data sets employed: red squares
stand for $\beta=4.2$, green ones for $\beta=4.05$ and blue for
$\beta=3.90$. Right (left) plot shows estimates for momenta above
(below) 10 $GeV^2$. The physical value (in $GeV$) of the momentum in $x$-axis
is obtained by applying the ratios of lattice sizes in tab.1
 and $a(3.9)^{-1} = 2.301 GeV$.}
\end{figure}

The deviations are found to be smaller than $5\%$ (see Tab. \ref{tab:scaling}), 
as in the quenched case. This deviation
might be a signal of discretization errors still present at these $\beta$'s.
Another source of discrepancy could be a possible dependence of results on the quark masses. Further efforts should be done in this sence.

\subsection{$\Lambda_{\overline{MS}}$ and $\VEV{A^2}$ condensate}

\begin{figure}[h!]
\begin{center}
\includegraphics[width=10cm]{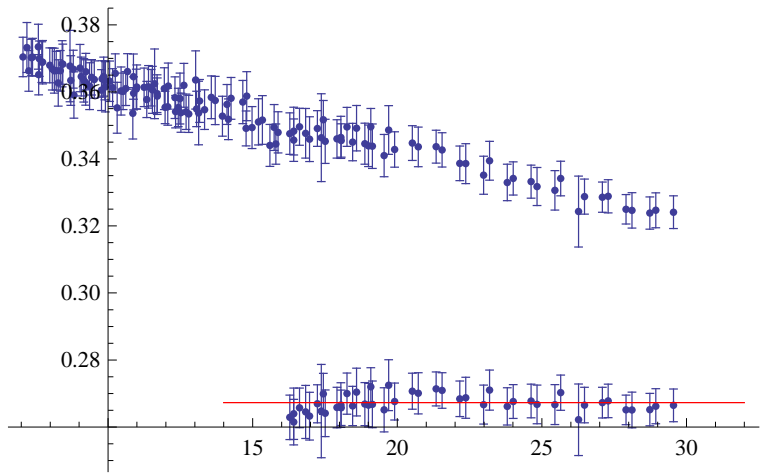}
\end{center}
\caption{\small $\Lambda_{\overline{\rm MS}}$ derived from fitting
the lattice value of $\alpha_T$ with the perturbative+OPE
prediction, in terms of the momentum where $\alpha_T$ is estimated
from the lattice, as described in ref.~\cite{Boucaud:2008gn}.}
\label{fig:lambda}
\end{figure}

The value of $\Lambda_{\msbar}$ can be obtained by inverting 
(\ref{betainvert}) for the lattice values of $\alpha_T$ obtained
from the lattice for each momentum. When done (figure \ref{fig:lambda})
the values of $\Lambda_{\msbar}$ obtained have a strong dependence on the 
momentum, showing the presence of some non-perturbative effects not taken
into account in (\ref{betainvert}). The values of $\Lambda_{\msbar}$ are 
around $320-360\mev$, much higher than other estimations.

The first non-perturbative correction that does appear un Landau gauge is the 
$\vev{A^2}$ gluon condensate, whose effects on the running coupling are included
in (\ref{alphaNP}). The values of $\Lambda_{\msbar}$ and $\vev{A^2}$ can be simultaneously 
fixed from lattice data using, for example, the ``plateau'' method, 
shown in \cite{Boucaud:2008gn}. It consist in varying the value of the
condensate to look for a ``plateau'' in $\Lambda_{\msbar}$ over a given momentum
window.

In fig.~\ref{fig:lambda}, we also
plot $\Lambda_{\overline{\rm MS}}$ derived from confronting the
lattice value of $\alpha_T$ with the perturbative+OPE prediction,
in terms of the momentum where $\alpha_T$ is estimated from the
lattice. The application of the  ``plateau'' method allows us to 
get as a best estimate:
\beq \Lambda_{\overline{\rm MS}} \ = \ 267 \pm 11 {\rm MeV} \ ;
\eeq
where again the error takes into account no systematic effect.
This result is in good agreement with other estimations in 
litteraure~\cite{Gockeler:2005rv}-~\cite{DellaMorte:2004bc}.
The value of the  $\VEV{A^2}$ obtained is
\beq
g_T^2\VEV{A^2}_{R,\mu} = 9.6 \pm 0.6 \gev^2
\eeq
which shows a significant increase respect to previous
quenched estimates~\cite{Boucaud:2008gn}.

\section{Conclusions and outlooks}

We calculated the running coupling in the Taylor scheme with $N_f=2$ 
flavours of dynamical quarks. We found that the matching of the 
results obtained for different $\beta$'s allows to compute the 
ratio of lattice spacings, with a deviation with respect to the string
tension always smaller than $5\%$.

By comparing the lattice result with the expectation coming from 
perturbation theory, we found the need for a dimension-two gluon 
condensate associated to a non-perturbative power correction. 
Including this term allows for an agreement between lattice and 
continuous formulae and then the extraction of the scale 
$\Lambda_{\msbar}^{N_f=2}$. Our result is in agreement with previous 
determinations.

The application of this method is straightforward for
a higher number of quark flavours and might be used in 
forthcoming $N_f=2+1+1$ lattice simulations.

As an outlook, we are interested in checking the mass-dependence of 
our results. In particular two effects are to be expected. The first one, 
at the level of the calibration, could show a dependence of the lattice spacing 
both on $\beta$ and $\mu$. In any case this should not affect our results.
The second one could be the effect of the mass on the coupling, which seems to 
be rouled out because of the good overlap of the coupling already observed 
at different $\mu$'s.

\subsection{Acknowledgements} We thank the IN2P3 Computing Center 
(Lyon) where our simulations have been done.

\end{document}